\documentclass [12pt] {article}

\catcode`\@=11
\@addtoreset{footnote}{section}
\catcode`\@=12

\usepackage{amssymb}
\usepackage {latexsym}

\newcommand {\id} {\mathbb I}

\newcommand {\pa} {\partial}
\newcommand {\pb} {\bar \pa}

\newcommand {\vary} {\delta}

\newcommand {\union} {\cup}

\newcommand {\beq} {\begin {equation}}
\newcommand {\eeq} {\end {equation}}

\newcommand {\beqn} {\begin {displaymath}}
\newcommand {\eeqn} {\end {displaymath}}

\newcommand {\beqar} {\begin {eqnarray}}
\newcommand {\eeqar} {\end {eqnarray}}

\newcommand {\beqarn} {\begin {eqnarray*}}
\newcommand {\eeqarn} {\end {eqnarray*}}

\newcommand {\nono} {\nonumber \\ {}}

\newcommand {\bary} {\begin {array}}
\newcommand {\eary} {\end {array}}

\newcommand {\half} {\frac 1 2}

\newcommand	{\come} {\;\;\;\;}

\newcommand {\eqr} [1]  {{(eq. \ref {eq:#1})}}
\newcommand {\Eqr} [1]  {{(Eq. \ref {eq:#1})}}

\newcommand {\secr} [1] {\S \ref{sec:#1}}

\newcommand {\ignoretext} [1] {}

\newcommand {\RR} {{\mathbb  R}}

\newcommand{\brac}[1]{\left\{{#1}\right\}}	\newcommand{\brak}[1]{\left[{#1}\right]}	\newcommand{\bracP}[1]{\left({#1}\right)}

\newcommand{\set}[1] {\brac {{#1}}}

\newcommand{\inv}[1] {{#1}^{-1}}

\newcommand{\transpose}[1] {{#1}^{T}}

\newcommand{\bvbtm} {\begin{verbatim}}
\newcommand{\evbtm} {\end{verbatim}}

\newcommand {\restrict}[2][] {{\left. #2 \right|_{#1}}}

\newcommand {\R} {R}

\newcommand {\cpo} {\circ}

\newcommand {\image} {{\mbox {Im}}}

\newcommand {\rank}	{{\mbox {rank}\,}}

\newcommand {\bdry} {\pa}

\newcommand {\intersect} {\cap}

\newcommand {\tensorUD}[3] {{{#1}^{#2}}_{#3}}

\newcommand {\mqn} {{\mathcal W}}

\newcommand {\OOY} {Ooguri:1996ck}
\newcommand {\BBMOOY} {Becker:1996ay}
\newcommand {\BMM} {Breckenridge:1996tt}
\newcommand {\WInstanton} {Witten:1995gx}
\newcommand {\WBSSP} {Witten:1995im}
\newcommand {\DLP} {Dai:1989ua}
\newcommand {\Leigh} {Leigh:1989jq}
\newcommand {\Townsend} {Townsend:1995af}
\newcommand {\GHM} {Green:1996dd}
\newcommand {\Yin} {Yin:2002wz}
\newcommand {\CY} {Cheung:1998az}
\newcommand {\nextY} {Yin:2appear}

\newcommand {\Lanczos} {Lanczos:1986VPM}
\newcommand {\GSW} {Green:1987sp}
\newcommand {\Ishibashi} {Ishibashi:1989kg}
\newcommand {\Maupertuis} {Maupertuis:1749EC}

\usepackage{nohyperref}

\makeatletter
\@addtoreset{equation}{section}

\renewcommand {\eqr} [1]  {{eq. (\ref {eq:#1})}}
\renewcommand {\Eqr} [1]  {{Eq. (\ref {eq:#1})}}

\newcommand{\thetitle} {
THE PRINCIPLE OF LEAST ACTION \\ AND \\ THE GEOMETRIC BASIS OF D-BRANES}

\newcommand{\theauthorlist}{Zheng YIN\footnote{yinzheng at ustc point edu dot cn}}

\newcommand{\theaddress} {
Center for Mathematics and Theoretical Physics\\
Shanghai Institute for Advanced Studies\\
University of Science and Technology of China\\
99 Xiupu Rd, Nanhui, Shanghai, China, 201315
}

\newcommand{\theabstract} {
We analyze thoroughly the boundary conditions allowed in classical
non-linear sigma models and derive from first principle the
corresponding geometric objects, i.e. D-branes. In addition to giving
classical D-branes an intrinsic and geometric foundation, D-branes 
in nontrivial H flux and D-branes embedded within D-branes are precisely 
defined.  A well known topological condition on D-branes is replaced.
}

\newcommand{\thepreprintnumbers}{
SIAS-CMTP-05-3 \cr
hep-th/0601160
}

\begin {document}

\begin{titlepage}
\hfill
\vbox{ 
\halign{#\hfil     \cr   
\thepreprintnumbers \cr
           } 
      }  
%           \hfill \verb=pre-rel-$Id: arxiv_1-n=0_gen.tex,v 1.3 2006/01/22 18:09:50 yin Exp $=
\vspace*{20mm}
\begin{center}

{\Large {\bf  \thetitle}\\}
\vspace*{15mm}
{\theauthorlist}

\vspace*{5mm}
{\it {\theaddress}}\\

\vspace{10mm}

\end{center}

\begin{abstract}
\theabstract
\end{abstract}

\vspace*{15mm}
\flushleft December 2005

\end{titlepage}

%\verb CVS $Id: arxiv_1-n=0_gen.tex,v 1.3 2006/01/22 18:09:50 yin Exp $

%CFFR: Comment out before submission

% The content starts below.

\section {Introduction and conclusion}

There have been a great deal of works involving D-branes. Many connect
different formalisms and limits using dualities and extrapolate from
different branches of mathematics, relating to a diverse range of
topics. This reflects the richness of string theory in general but
complicates our understanding of D-branes because it is often difficult
to pin down a supposed feature of D-brane to any well defined set of
assumptions.  Very little if any had been done for the foundation of
D-branes theory.

The purpose of this paper is to fill this void and carry out a systematic
analysis from first principle.  Our object of study is the classical
non-linear sigma model; our point of departure is the 
principle of least action \cite{\Maupertuis}\footnote 
	{See \cite {\Lanczos} for a modern introduction}.  
Neumann boundary condition for Type I string in flat
spacetime has been formulated this way in textbooks\cite {\GSW}, and Dirichlet
boundary condition was known to be compatible with it \cite {\Leigh}. 
But until now a thorough and general treatment emphasizing logical coherence 
rather than examples and applications had been absent. We believe such an
undertaking is worthwhile given the important roles D-branes currently
play in theoretical physics.
In this paper we examine the boundary
variational problem carefully and thoroughly and solve it in 
well defined generality.
D-brane background is then understood as the collection of data completely
characterizing a solution.

Since our work is foundational rather than applied in nature, it is
perhaps surprising that in addition to putting some known aspects of
D-branes on a firm ground and explicating their domain of validity,
we have also obtained new and even unexpected
results.  Thanks to the comprehensiveness of our analysis we uncovered
solutions not known before.  As one of the more striking
consequences, a standard argument for a topological
condition on D-branes, that the pull-back of the NS 3-form $H$ to it
must have vanishing cohomology class, is invalidated.  That argument was
based on the premise that a D-brane must have a $2$-form flux $F$ defined on
it whose exterior differential agrees with the pull-back of $H$.  We
have shown instead that the data defining a D-brane is a submanifold $S$ and a certain 
tensor $\R$ defined over it, while $F$ does not have to be well defined
over $S$.  $\R$ has a far more intricate relationship with $H$ that does
not lead to an obvious obstruction for the submanifold per se.  
We give a counterexample of a D-brane that fails the old condition and derive 
the correct relation between $H$ and backgrounds on D-branes replacing it.
The  exact nature of D-branes in the
presence of $H$ field had always been a puzzle, which is finally solved
in this work.  Also interesting is the appearance of the
embedding of one D-brane within another D-brane as a solution.

While this paper is oriented toward the physics side of D-branes, in the
course of this investigation we have also obtained several mathematical results
on the geometry of D-branes which will appear in \cite {\nextY} 
in an appropriately abstract setting.  Some of them are announced in this paper.

Our work diverges from other worldsheet studies of D-branes in the
literature on the issues of conformal invariance and classical
vs. quantum.  Our treatment is strictly classical.  Although a quantum
treatment would certainly be useful, it is usually unfeasible.  Much can
already be learned from a classical analysis as a limit of the quantum
theory.  We note also our
treatment is directly relevant for the quantum theory because path
integral quantization is also based on an action principle.

Conformal invariance is never invoked in this work, but it will be
evident that the boundary conditions we derive are automatically compatible with
(classical) conformal symmetry, namely breaking it by half.  
Of course, quantum fluctuation usually
renders scale invariance anomalous.  As a result, quantum
treatment of conformal invariant boundary condition typically employs
boundary conformal field theory \cite {\Ishibashi}.  
While this allows one to consider
D-branes in any conformal field theory, including those without
a spacetime interpretation or a Lagrangian formulation, it often obscures
the fundamental geometric nature of D-branes.  On the other hand, while
our approach does not ensure conformal invariance after quantization, it
extends the geometric notion of D-brane far beyond conform field
theories.  In fact, our analysis is applicable without change when the
bulk Lagrangian in \eqr{sigmaAction} is modified by anything not
involving $X'$. Another systematic approach to the
problem of (super-)conformal invariant boundary condition considers the
geometry of (super-)boundaries on (super-)worldsheet \cite {\Yin}.

After D-branes were first constructed as flat extended objects in flat
spacetime on which an open string ends \cite{\DLP}, one expects this
open string worldsheet approach to be generalized to manifolds. Modern
treatment began with \cite {\OOY} and \cite {\BBMOOY}, which established
the widely used worldsheet method for wrapping D-brane around an
arbitrary manifold in arbitrary spacetime. However, subsequent works had
not managed to really derive the standard D-brane boundary conditions
formulated in \cite {\OOY} from first principle.  We do this in section
two of this paper.  In section three we examine these conditions and
find a much more general set of D-brane solutions.  In section four we
present two examples of these new solutions.

\section {Variation}

\subsection {On $\bdry \Sigma$}		\label {sec:variation}

The 2d classical sigma model on a boundary is described by the action 
\beq	\label {eq:sigmaAction}
	S = \int_{\Sigma} \bracP {\half G(\dot X, \dot X) - \half G(X', X') 
	+ B(\dot X, X')}
	+ \int_{\bdry \Sigma} i_{\dot X} A,
\eeq
Here $M$ is the target space manifold and $\Sigma$ is a smooth 2d
manifold with boundary $\bdry \Sigma$. $X: \Sigma \to M$.  Locally
$\Sigma$ is parameterized by coordinates $t$ and $\sigma$.  $\dot
X$ and $X'$ are derivatives with respect to $t$ and $\sigma$
respectively. $G$ is a
nondegenerate metric defined on $M$.  $B$ is a locally defined $2$ form
whose exterior derivative is a well defined $3$-form $H$ on $M$. 
$i_{\dot X}$ is the insertion operator of $\dot X$ acting on covariant
tensors of arbitrary rank.

The integral over the boundary is the standard boundary extension of the
$\sigma$ model action,
where $A$ is a 1-form whose exact nature we will address presently.
It is worth noting that in addition to being explicitly
geometric the action \eqr{sigmaAction} also possesses conformal symmetry in
$2$ dimensions.

Because the worldsheet has boundary, the variation of $S$ in the interior
contributes a boundary term in addition to that from the boundary
action.  The solution to the variational problem therefore consists of the
equation of motions in the interior of $\Sigma$ and a set of boundary
conditions on $\bdry \Sigma$.  In this paper we are interested in
the latter.  Each connected component of the boundary gives
rise to an independent condition but they are all of the same structure.
So it suffices to consider one such boundary, denoted by $\bdry_1 \Sigma$.
We shall call a boundary condition on such a component a \emph
{D-brane boundary condition}.  The data characterizing a D-brane
boundary condition is called a \emph {D-brane background solution} (to
the boundary variational problem) of the $\sigma$-model.  Variation
principle requires
\beq	\label {eq:variationCondition}
	\restrict[\bdry_1 \Sigma] 
		{\bracP {G \bracP {\vary X, X'} - i_{\vary X} i_{\dot X} e^*B 
		+\vary{\bracP {i_{\dot X} A}}}} = 0.
\eeq
This is \emph{the} boundary condition required by the principle of least
action for \eqr{sigmaAction}.
We consider here solving it by first imposing a condition on
\beq	\label {eq:XBC}
	\restrict[\bdry_1 \Sigma] {\vary X} \come.
\eeq
$\vary X$ takes value in $\restrict[X] {TM}$.  Because $M$ is an arbitrary
manifold, it is not possible to restrict $\vary X$ without explicitly
referring to $X$.  The only way to restrict $\restrict[\bdry_1 \Sigma]
{X}$ is to restrict it to
a subset $S$ of $M$.  As far as the
variational problem is concerned, we should treat separate components of 
$S$ separately because variation cannot traverse across separated components.
Therefore we may assume $S$ is a connected set.  
But this is not yet
enough.  We expect that the end points should be able to move
continuously within $S$.  Furthermore we want to perform standard
calculus on $S$ so that we can derive equation of motion at the
boundary.  If we assume nonsingularity, i.e. every point in $S$ has a
neighborhood in $S$ that is homeomorphic to the same set, that set must
be homeomorphic to $\RR^n$ for some positive integer $n \leq d$.  It
then follows that $S$ is a $n$ dimensional submanifold of
$M$.  Note
that we have reached this conclusion through very general consideration.  
This is the first piece of data of a D-brane background solution and we shall 
say the D-brane is wrapping the submanifold $S$ if it meets certain
conditions stated below.

So far we have only considered how to restrict $\restrict[\bdry_1 \Sigma] {X}$, 
but we really want to restrict $\restrict[\bdry_1 \Sigma] {\vary X}$.  
Consider a particular point $t \in \bdry_1 \Sigma$ and let $x= X(t)$. 
Denote by ${\mqn(x)}$ the subspace of $\restrict[x] {TM}$ to which
$\vary X (t)$ must belong.  Obviously $\mqn(x)
\subset \restrict[x] {TS}$, the subspace of $\restrict[x] {TM}$ tangent
to $S$. It is tempting to think $\mqn(x)$ must coincide with
$\restrict[x] {TS}$.  In fact in previous works on the subject 
this had always been the implicit assumption.  We will for the
moment again make this assumption but revisit it in the next section.
Clearly $\dot X$ should be subjected to the same restriction as $\vary X$.

Given the above, it is now clear that $A$ exists only on $S$.  And on $S$ 
it is only locally defined but $F = dA + e^* B$ is well defined over $S$,
where $e^*$ is the pull-back map from the forms on $M$ to those on $S$.
In other words, it is gauge invariant \cite{\WBSSP},
because \eqr {sigmaAction} is gauge invariant under $B \to B + d
\Lambda$, and $A \to A - e^* \Lambda$.  It is
also $F$ rather than $A$ which enters the boundary condition: 
\eqr{variationCondition} can now be rewritten as
\beq \label {eq:variationConditionF}
	\restrict[\bdry_1 \Sigma] 
		{\bracP {G \bracP {\vary X, X'} - i_{\vary X} i_{\dot X} F }} = 0.
\eeq

Now it follows that 
\beq	\label {eq:HCond}
	dF = e^*H,
\eeq
This is the standard argument leading to the well known topological
condition on D-branes:
\beq	\label {eq:TopoCond}
	[e^*H] = 0
\eeq
in the De Rham cohomology of $S$.  

Together with $\restrict[\bdry_1 \Sigma] X \in S$ this is a complete set
of D-brane boundary conditions for the boundary variational problem
on $\bdry_1 \Sigma$. In fact all D-brane boundary conditions for
$\sigma$-model considered in the literature up to now are of this form,
corresponding to a D-brane wrapping the submanifold $S$ with flux $F$.
In the next section we shall show that a lot had been missed.

\subsection {Boundary conditions}	\label {sec:BC}

Let $S$ be an $n$ dimensional submanifold of $M$.  Consider a particular
map $X$, a point\footnote
	{$A \subset B$ means $A$ is a subset of $B$ while $A
	\subsetneq B$ means $A$ is proper subset of $B$.} 
$x \in X (\bdry_1 \Sigma) \subset S$, and a coordinate
patch $(U, x)$ in $M$ containing $x$.  With an abuse of notation we use
$x$ to also denote the coordination map $U \to \RR^d$.  We may assume
that the first $n$ coordinates also parameterize $S \intersect U$ and
denote them by $x^A, x^B, \ldots$; the rest are denoted by
$x^a, x^b \ldots$.  When no distinction between those two groups is desired,
greek indices $x^\mu, x^\nu, \ldots$ will be used.
We will also choose the coordinates on the worldsheet so that $t$ parameterizes 
$\bdry_1 \Sigma$.

With these notation in mind, \eqr {variationConditionF} yields\footnote
	{Summation over repeated indices is implicit unless
	stated otherwise.}
\beqar	\label {eq:BCtwoPart}
	F_{AB} \dot X^B + G_{A\mu} X'^\mu &=& 0; \nono
	\dot X^a &=& 0.
\eeqar
The second equation stems from the restriction $\restrict[\bdry_1
\Sigma] {X} \in S$.

Now use light-cone derivatives $\pa = \pa_0 +
\pa_1$, $\pb = \pa_0 - \pa_1$, one finds that the following conditions hold
on $\bdry_1 \Sigma$:
\beqar	\label {eq:BCspecCoord}
	(G_1 - F) \pb X &=& 	(G_1 + F) \pa X + 2 G_2 \pa X \nono
	\pb X^a &=& - \pa X^a ,
\eeqar
where $G_1 = \bracP{G_{AB}}$, $G_2 = \bracP{G_{Ab}}$.
We shall assume that ${(G_1-F)}$ is invertible, which is
guaranteed if $G$ is positive definite.  Then \eqr{BCspecCoord} can
be rewritten as
\beq
	\pb X^\mu = \tensorUD R \mu \nu \pa X^\nu,
\eeq
where $R$ is given by 
\beq	\label {eq:RinGF}
	\bracP {
	\bary{cc}
	\inv{(G_1-F)} (G_1+F) &		2 \inv {(G_1-F)} G_2 \\
	0	&	- \id
	\eary
	},
\eeq
Note that $\R$ is nondegenerate and like $F$ has $S$ as its domain.

Although we have written $R$ explicitly in components, it can be also be
written more abstractly \cite {\nextY}.  Thus $R$ is a well defined
section of the pull-back of the bundle of $(1,1)$ tensors on $M$ to $S$.
An important property of $R$ is \footnote
	{Note this is precisely the requirement for the boundary condition to
	preserve half of the  conformal invariance \cite {\OOY}.}
\beq	\label {eq:RGcompatibility}
	\transpose {R} G R = G
\eeq
if the pull-back of $G$ on $S$ is nondegenerate.
Here we give a proof in explicit components.  
If $G_1$ is nondegenerate,
$\mqn(x)$ has an orthogonal complement with respect to $G$.  So $\forall
x \in S$, one can find some coordinate patch containing $x$ so that $G_2$
vanishes at $x$.
There both $\R$ and
$G$ are block diagonal, as are the two sides of \eqr{RGcompatibility}. The
lower right block of the equation holds trivially.  For the upper left
block, consider\footnote
	{In accordance with custom, $G^{AC} \equiv (\inv{G})^{AC}$.}
$\tensorUD {\beta} AD \equiv G^{AC} B_{CD}$.  It satisfies $\transpose
{\beta} G = - G \beta$.  Then $(\id \pm \transpose {\beta}) G = - G (\id \mp
\beta)$.  The upper left block of $R$ is $\inv{(1-\beta)}(1+\beta)$.  
\Eqr {RGcompatibility} then follows.

The following is therefore also a complete set of boundary condition for
the variation problem of \eqr{sigmaAction}.
\beqar	\label {eq:BCTwo}
	\restrict[\bdry_1 \Sigma] X &\in& S; \nono
	R \, \pa X &=& \pb X.
\eeqar

\section {Analysis}

Because we have derived the second of \eqr{BCTwo} from \eqr{BCtwoPart}, the
two might appear to be equivalent, between which one can choose either
$R$ or $F$ as characterizing the background on the D-brane.   But this is
not quite right.  The $R$ obtained from the $F$ in the last section
is subject to a constraint: the multiplicity of its eigenvalue at $-1$ is
always $d-n$. Although this might appear
natural at first, it does not seem strictly necessary.  If we relax the
condition on $R$ and still find a boundary condition, it will be outside
the domain of validity of \eqr{BCtwoPart}.  In this section we examine
this possibility thoroughly.

\subsection {D-branes old and new}

Consider a pair $(S, R)$, where $S$ is a $n$ dimensional submanifold $M$
and $R$ is a section of the pull-back of the bundle of fiberwise
endomorphisms over $TM$ to $S$.  We require \eqr
{RGcompatibility}, which implies that $R$ is
diagonalizable. Define $P = \R + 1$.  Consider a neighborhood $N$ of a
point $x \in S$. It is clear that $\dot X \in \mqn \equiv \image P$.  For this to
be consistent with the first of \eqr{BCTwo} we must require $\mqn(x) \subset
\restrict[x] {TS}$ so $\rank P \leq n$.  The known cases correspond to
$\rank P = n$ and $\mqn = \restrict[x] {TS}$.  This still leaves the
possibility of $\mqn \subsetneq \restrict[x] {TS}$.

Suppose $\rank P = n - k$.  So we may diagonalize $R$ at $x$ with $k$
eigenvectors with eigenvalue $-1$.  Then 
$R$ is of the form \eqr {RinGF}, with the lower right block of dimension
$k$.  This can be shown by reversing the steps below \eqr {RGcompatibility}.
$k$ is clearly upper semi-continuous on $S$, i.e. it cannot
increase suddenly.  So without loss of generality we may assume $k$ does
not increase over $N$.  Of course $k$ may very well decrease in $N$. 
Consider the connected component containing $x$ of the set $N_k \equiv
\set {x \in N \, | \, \rank \mqn(x) = n - k}$.  In light of the consideration in 
\secr {variation}, this is a valid boundary condition if
$N_k$ is itself a submanifold of codimension $k$ and $\forall x \in N_k$, 
$\mqn(x) = TN_k$.

The case $\rank P < n$ is entirely new, because it is not translatable
to the pair $(S, F)$ describing D-brane boundary conditions with a
background gauge field.  However, it has a very intuitive interpretation
both physically
and mathematically.  If the submanifold $N_k$ exists, on the basis of
our early analysis it can and must be interpreted on its
own as a neighborhood of a $n-k$ dimensional (connected) D-brane, which we call
$S_k$.  Of course a given D-brane $S$ with background $R$ may
contain several such D-branes.  Here we just consider one of
them.  The boundary conditions on $S_k$ are
prescribed by the pair $(S_k, \restrict[S_k] R)$ which on its own
provide a complete set of boundary conditions for the variational problem
of \eqr{sigmaAction}.  However, it is also a part of the boundary
conditions specified by $(S, R)$ in the precise mathematical sense just
described.
Therefore we have a D-brane $S_k$ with background $\restrict[S_k] R$
embedded in the bigger D-brane $S$ with background $R$.  Furthermore, by
reapplying the analysis above, one sees that
$S_k$ itself may contain lower dimensional D-branes, say some
${(S_k)}_l$ in an obvious extrapolation of previous notation.  The latter
can also be interpreted directly as $S_{k+l}$.  Therefore we can
consider nested structure of D-branes within D-branes as submanifolds
contained within submanifolds with background $R$ defined on them through
restriction.  The submanifold structure is arbitrary but entirely
determined by the background $R$.  Define 
$S^\union_k \equiv \set {x \in S \,|\, \rank
\mqn (x) \leq n - k}$.  $S^\union_k$ is the union of all possible $S_k$'s.  
We clearly have the following filtration:
\beq	\label {DbraneFiltration}
	S^\union_n \subset \ldots S^\union_1 \subset S^\union_0 = S.
\eeq

There is an analytical constraint on the filtration.  $R$ is nondegenerate, so the
sign of its determinant cannot vary continuously over $S$.  $R$ is real, so its
eigenvalues are either real or in complex conjugate pairs.  $R$ is
orthogonal, so they are either $e^{\pm i\theta}$ or $\pm 1$.  It
follows then the parity of the multiplicity of the eigenvalue $-1$ of $R$ cannot vary
continuously over $S$.  When considering continuously varying $R$, we need 
to consider only $S_k$ for
even $k$.  In such cases, the embedded branes always have even
codimensions.

\subsection {Physical interpretation}

While we have tried to avoid excessive abstraction, so far our emphasis 
is analytical.  Yet what we found has a very direct and clear
physical interpretation, even though the context is a little surprising.
D-branes, after all, are dynamical extended objects in superstring theory. It
is well known that in superstring theory, a Dp-brane can carry RR
charges for the D(p-2)-brane \cite {\Townsend}, thus forming a bound
state.  This structure can be nested, and the codimensions may be any even 
number bounded by $p$.  This is precisely what we have found!  However, there
are some very intriguing aspects of our results.  The identification of embedded
branes in superstring theory required various dualities \cite
{\WInstanton}, \cite {\Townsend}, \cite {\BMM} and the relation between
their RR charge and the gauge field on the D-branes was understood from
anomaly cancellation for the D-brane worldvolume theory, which contains
chiral fermions \cite {\GHM}\footnote
	{Note: while it was correct in using anomaly inflow to deduce the RR
	coupling of D-branes, \cite {\GHM} missed an important part of brane
	anomalies as well as a key subtlety in the action of RR fields. See
	\cite {\CY} for a detailed general treatment.  The differences have
	profound implications for the physics and mathematics of
	D-branes.}\cite {\CY}.
All those arguments are only applicable to superstring theories. The analysis in
this paper can be generalized to the supersymmetric $sigma$ model \cite
{\nextY}, and the same phenomenon obviously persists  So our results agree well
with the known results and provides the first \emph{exact} description
of a localized embedded D-brane.  Because our analysis is carried out
from first principle and independent of any duality argument, it can be
seen as providing indirect support
for the validity of several string dualities.  What is the most striking,
however, is that the domain of validity of our analysis is much more
general than superstring theories and applicable \emph {directly} to any
$\sigma$-model type action as mentioned in the introduction, suggesting
that this phenomenon is universal.

It has been long observed that the background $F$ interpolates
between Neumann and Dirichlet boundary conditions.  So if the
component of $F$ over a plane is taken to infinity then in some
sense these directions become Dirichlet because they will have $-1$
eigenvalue, indicating a ``smaller'' brane.  The hierarchy of
$S^\union_k$ we find above finally give this
intuitive idea a concrete and mathematically precise realization.

\subsection {General solutions}	\label {sec:mannequin}

It turns out that there are a lot more new D-brane background solutions
besides the embedded branes discussed above.  Rather than
examining them case by case, we take a general ground and look for
the \emph {most} general of D-brane boundary conditions for 
the sigma model.  This entails finding the
most general form of the boundary action, because a variational problem
is determined by an action.

We now describe a most natural generalization of the boundary part of
\eqr{sigmaAction} that we have discovered.  We will emphasize the
intuitive and conceptual aspects relevant for the physical problem and a
physicist's viewpoint, while reserving the detail of the formal
development and generalization for \cite {\nextY}. Let $\mqn$ be a general
distribution on $S$ such that
every open set in $S$ contains a point $x$ at which $\mqn(x) =
\restrict[x] {TS}$. We say $\mqn$ is a \emph {mannequin} for $S$. By
this definition, there may be a nonempty set of points where $\rank
\mqn$ is not at a local maximum, but its complement must be dense in
$S$.

Consider now the dual $\mqn^*$ in $T^*M$: $\forall x \in S$, $\mqn^*(x) =
\bracP{\mqn(x)}^*$.  Sections of $p$-th exterior powers of $\mqn^*$ will
be called $\mqn$ $p$-forms.  The inclusion of $\mqn$ in $TS$ induces a
pull-back map $e_\mqn^*$ from p-forms on $S$ to $\mqn$ p-forms.  We require
$\mqn$ to be involutive, i.e. its sections are closed under Lie bracket.  
One can therefore define a complex of $\mqn$
p-forms in the usual fashion with the following exterior differential.
\beqar
	&& d \omega (V_0, \ldots, V_n) 
	\equiv \sum_{i=0}^n (-1)^i \{ L_{V_i} \;
		\omega ( {\ldots, V_{i-1}, V_{i+1}, \ldots } ) \nono
		&& - \sum_{j = i+1}^n 
			\omega ( { \ldots, V_{i-1}, V_{i+1}, \ldots, V_{j-1},
			\brak {V_i, V_j}, V_{j+1}, \ldots } ) \}.
\eeqar
Then $e_\mqn^*$ is a chain map.

We are now ready to write down the general boundary action.  Given a
submanifold $S$ of $M$, a mannequin $\mqn$ of $S$, and a locally defined 
$\mqn$ 1-form $A$, consider the following replacement for the boundary
part in \eqr {sigmaAction}:
\beq
	\int_{\bdry_1 \Sigma} i_{\dot X} A.
\eeq
The total action is gauge invariant under $B \to B + d \Lambda$, and 
$A \to A - e_\mqn^* \cpo e^* \Lambda$.  Therefore $F \equiv dA +
e_\mqn^* \cpo e^* B$ is a well defined gauge invariant $\mqn$ 2-form.  
With the condition
\beqar
	\restrict[\bdry_1 \Sigma] X &\in& S; \nono
	\restrict[\bdry_1 \Sigma] {\vary X}\,,\, \restrict[\bdry_1 \Sigma] {\dot X}
		&\in& \mqn(X)
\eeqar
variational principle then leads to the following equation
\beq
	\restrict[\bdry_1 \Sigma] {e_\mqn^* \cpo e^* \cpo i_{X'} G - i_{\dot X} F} = 0.
\eeq
By the same
analysis as in \secr{BC}, we obtain again a matrix $R$ relating $\pb X$
to $\pa X$ when the nondegeneracy conditions stated there holds.  
This shows that pair $(\mqn, F)$ is in fact in 1-1 correspondence with
$R$ under those conditions.  A more abstract derivation is also
possible\cite {\nextY}.

As before it is $F$ rather than $A$ that enters in the boundary condition. 
The condition for $F$ is clearly
\beq	\label {eq:myHCond}
	d F = e_\mqn^* \cpo e^* H
\eeq
This replaces \eqr{HCond}.  This change however is not a mere
generalization but instead carries very different meaning and significant
ramification. \Eqr {TopoCond} is a topological condition on $H$ and the
submanifold $S$, and when $H$ has nontrivial cohomology class in $M$,
there are some submanifolds $S$ which are
excluded by this condition.  By contrast, \eqr {myHCond} depends on the
choice of an involutive mannequin, which seems quite arbitrary.  Even if a
submanifold $S$ is excluded by \eqr{TopoCond}, as long as we can find an
involutive mannequin $\mqn$ \emph {compatible} with $H$ in the sense that
\beq	\label {eq:myTopoCond}
	[e_\mqn^* \cpo e^* H] = 0
\eeq
in the $d$-cohomology of $\mqn$ forms, our analysis shows that a D-brane
can still wrap around $S$.  In \secr {SThreeExample} we will present an
instance of a submanifold which \eqr{TopoCond} says cannot be wrapped by
D-brane but which admits an exotic D-brane background solution of the
type that we have discovered.  Therefore \eqr {TopoCond} is invalidated.
Does
\eqr{myTopoCond} imply a more lenient condition on $S$ or simply removes
any? While we do not yet have a proof, based on observational
evidence we conjecture that any manifold can be wrapped by D-brane in
the context of classical $\sigma$-model.   Namely, for any given $H$ and
$S$, we conjecture there is some involutive mannequin for $S$ compatible
with $H$.

In summary, for a spacetime triple $(M, G, H)$, where $M$ is a
differential manifold, $G$ a nondegenerate metric on $M$, and $H$ a closed 
3-form on $M$.  A D-brane background solution consists of $(S, \mqn, F)$,
where $S$ is a submanifold in $M$, $\mqn$ an involutive mannequin for $S$, and $F$ a
$\mqn$ 2-form satisfying \eqr{myHCond}.  The mannequin $\mqn$ is said to
wear the D-brane on $S$ with background $F$.

\section {Examples}

In this section we consider two exotic examples of D-branes background 
solution not known before.

\subsection {D2-brane wrapping $S^2$ with a D0-brane embedded}

Consider a D2-brane having the topology of $S^2$.  We shall work in 
spherical coordinates and use the $O(3)$ symmetric metric
$ds^2 = d\theta^2 + \sin^2\theta \, d\phi^2$.  Consider the involutive
mannequin given by the linear hull of
\beqar
	\pa_\phi&,& \nono
	\cos \frac \theta 2 \sin \phi \; \pa_\theta 
	&+& \frac {\cos \theta \cos \phi} {2 \sin \frac \theta 2} \: \pa_\phi, \nono
	\cos \frac \theta 2 \cos \phi \; \pa_\theta 
	&-& \frac {\cos \theta \sin \phi} {2 \sin \frac \theta 2} \: \pa_\phi
\eeqar
everywhere and the background 
\beq
	F = 2 \sin^2 \frac \theta 2
	\bracP {\bary {rr} 
			0	& -1 \\
			1 & 0
		 \eary
		}
\eeq

This corresponds to the following $\R$ background
\beq
	\bracP { \bary {rr} 
		\cos \theta	& - \sin^2 \theta \\
		1 & \cos \theta
			 \eary
	}
\eeq
$R$ is smooth everywhere and develops two Dirichlet directions at the
south pole, indicating that a D0-brane is embedded.  $F$ is defined
everywhere but there.

\subsection {D-brane on $S^3$ with nontrivial $H$}	\label {sec:SThreeExample}

The simplest manifold which the supposed topological condition
\eqr{TopoCond} would have prohibited a D-brane from wrapping 
is an $S^3$ over which
there is a $H$ field of nonvanishing De Rham cohomology.

Employ the Hopf coordinates $\theta/2$, $\phi$, and $\rho$, so that the
volume form is $H = \sin \theta \, d\theta \, d\phi \, d\rho$ and $\rho$ is the
coordinate for the circle fiber.  Consider the 2-form 
\beq
	F = 2 \sin^2 {\theta/2} \, d\phi \, d\rho.
\eeq
It becomes singular at the south pole but everywhere else its
differential is the volume form.  This background $F$ is well defined for 
the involutive mannequin of $S^3$ that is everywhere the linear hull of
\beqar
	\pa_\phi, && \pa_\rho, \nono
	\cos \frac \theta 2 \sin \phi \; \pa_\theta 
	&+& \frac {\cos \theta \cos \phi} {2 \sin \frac \theta 2}\: \pa_\phi. \nono
	\cos \frac \theta 2 \cos \phi \; \pa_\theta 
	&-& \frac {\cos \theta \sin \phi} {2 \sin \frac \theta 2}\: \pa_\phi.
\eeqar
Therefore $H$ satisfy \eqr{myHCond} and hence D-brane can wraps this
$S^3$.

\newpage

\bibliographystyle{hepunsrt}

\bibliography{string-literature}

\end {document}